\documentclass[11pt]{article}

\usepackage[preprint]{acl}

\usepackage{times}
\usepackage{latexsym}
\usepackage{xcolor}

\usepackage[T1]{fontenc}

\usepackage[utf8]{inputenc}

\usepackage{microtype}

\usepackage{inconsolata}

\usepackage{graphicx}
\usepackage{amsmath}
\usepackage{amssymb}
\usepackage{amsfonts}
\usepackage{booktabs}
\usepackage{adjustbox}
\usepackage{multirow}
\usepackage{hyperref}

\usepackage{tabularx}
\usepackage{tikz}
\usepackage{subcaption}

\usepackage{todonotes}

\definecolor{mygreen}{RGB}{0,120,0}
\title{From Sounds to Scenes: A Benchmark for Evaluating \underline{C}ontext-Aware \underline{A}uditory \underline{S}cene \underline{U}nderstanding in Large Audio Language Models}

\author{
 \textbf{Pengfei Zhang\textsuperscript{1}},
 \textbf{Hoang H Nguyen\textsuperscript{2}},
 \textbf{Kazi Shaharair Sharif\textsuperscript{3}}
 \textbf{Yutong Song\textsuperscript{1}},\\
 \textbf{Wenjun Huang\textsuperscript{1}},
 \textbf{Henry Peng Zou\textsuperscript{2}},
 \textbf{Pinxin Liu},
 \textbf{Honghui Xu \textsuperscript{3}},
 \textbf{Amir M. Rahmani \textsuperscript{1}}
\\
\\
 \textsuperscript{1}University of California Irvine,
 \textsuperscript{2}University of Illinois Chicago,
 \textsuperscript{3}Kennesaw State University
}

\begin{document}
\maketitle

\begin{abstract}
Recent Large Audio Language Models (LALMs) have achieved remarkable progress in audio perceptual tasks across individual acoustic layers, including speech, sound, and music. However, existing benchmarks predominantly evaluate these layers in isolation, overlooking the complex contextual relationships that arise when multiple acoustic sources co-occur in real-world auditory scenes.
Real-world auditory interpretation requires \textbf{Context-Aware Auditory Scene Understanding (CASU)}: the ability to comprehend the holistic scene by integrating sound layers. 
To evaluate this capability, we introduce the CASU benchmark, which assesses whether Audio LLMs can interpret auditory scenes composed of speech, acoustic events (e.g., announcements), and background environments (e.g., traffic), and reason about the logical relationships between these layers.
We propose a scalable pipeline for constructing time-accurate, semi-synthetic audio streams by composing real-world scene sounds with synthetic speech. Building on this data, we design four tasks that probe scene understanding: contextual question answering, entity extraction from the scene, speaker role inference, and counterfactual reasoning where scene is manipulated.
Experiments across multiple LALMs demonstrate that effective auditory scene understanding requires integration over all auditory layers, rather than reliance on speech or sound alone, underscoring the necessity of CASU for advancing complex audio understanding in LALMs.
\end{abstract}

\begin{figure}[ht]
    \centering
    \includegraphics[width=\linewidth]{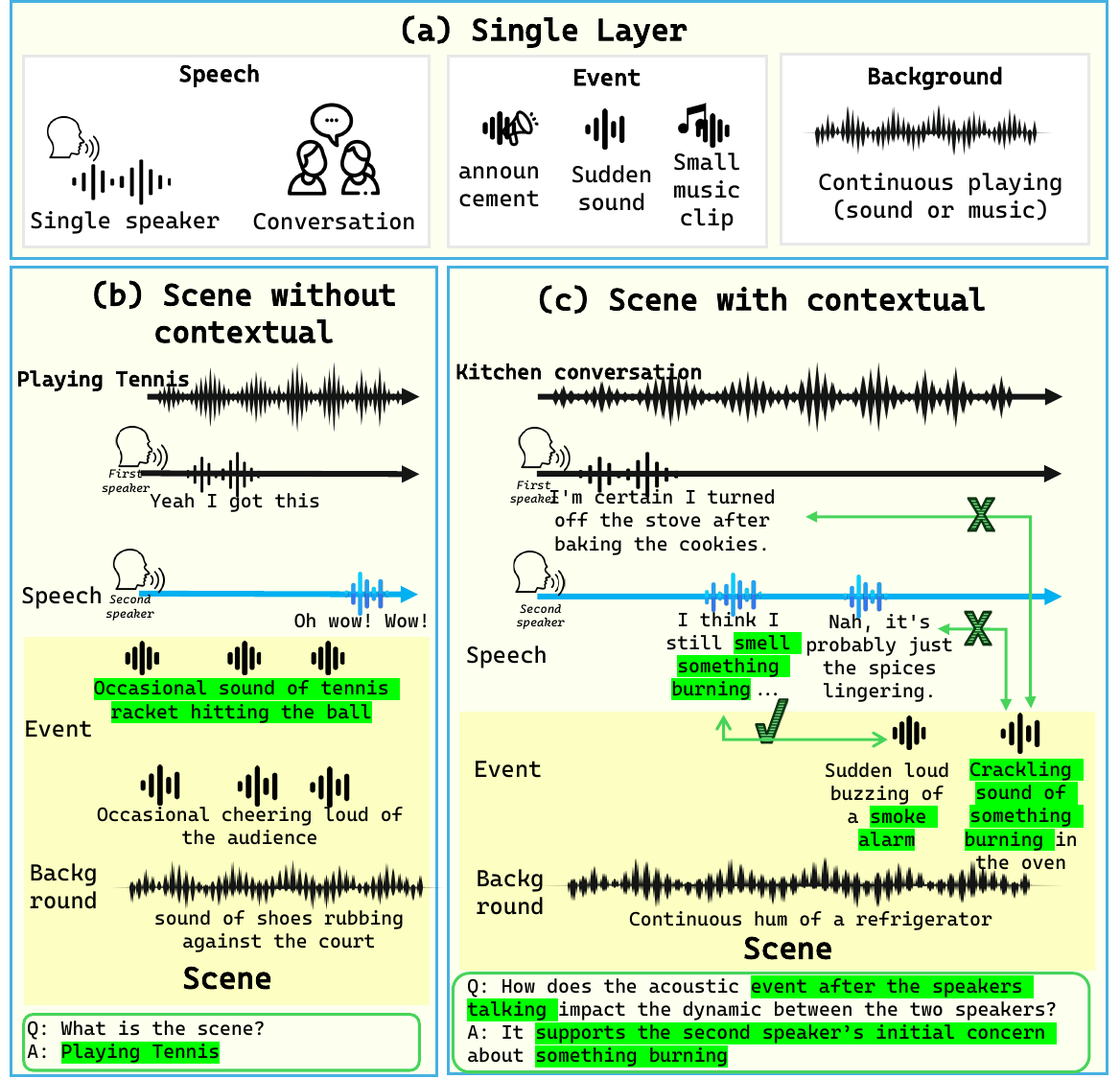} 
    \vspace{-0.6cm}
    \caption{\textbf{Comparison of auditory understanding paradigms across benchmarks.}
\textit{\textbf{(a) Single-layer evaluation}}, where prior work focuses on isolated acoustic layers, no cross-layer interactions. \textit{\textbf{(b) Mixed-layer scenes without contextual reasoning}}, introduced by recent benchmarks, where sounds of speech, events, and background co-occur, but contextual relationships are weakly expressed or not considered for answering questions. \textit{\textbf{(c) Context-Aware Auditory Scene Understanding (CASU)}}, where sounds of speech, events, and background jointly form a scene. In CASU, elements inside scene may confirm, contradict, or contextualize each other.
Green highlights indicate contextual relationships in the scene related to the question.}
    \label{fig:fig1}
\end{figure}

\section{Introduction}

Recent Large Audio-Language Models (LALMs) \cite{Qwen3-Omni,comanici2025gemini} have demonstrated strong performance across a range of audio perceptual tasks, including Automatic Speech Recognition, sound event detection, and audio captioning. These models effectively bridge acoustic signals and reasoning, but existing benchmarks primarily evaluate such capabilities at the level of isolated acoustic layers, i.e, speech, sound, or music. As illustrated in Figure 1(a), audio understanding is often framed as a single-layer perception task\cite{sakshi2024mmau,clark2018think}. Even recent benchmarks that incorporate mixed-layer audio inputs (Figure 1(b)) typically treat co-occurring speech, events, and background sounds as loosely coupled\cite{ma2025mmar,kumar2025mmau}, without explicitly requiring models to reason about their contextual relationships.

In contrast, human auditory understanding is inherently scene-level and context-aware, integrating speech with surrounding acoustic events and background sounds to infer higher-order meaning\cite{lu2015context}. Consider the spoken phrase "It's time to go." If this speech coincides with a school bell (a transient event), it implies a routine dismissal; yet, if accompanied by an emergency siren, the same phrase becomes a critical warning to evacuate\cite{fuentes2022urban}. 
Crucially, non-speech audio elements do not play fixed semantic roles: depending on context, they may function as irrelevant noise or impose decisive logical constraints on the interpretation of spoken content. Effective auditory understanding therefore requires models not only to recognize multiple sound sources, but also to reason about whether and how these layers interact\cite{du2025crab}. This gap motivates \textbf{\underline{C}ontext-Aware \underline{A}uditory \underline{S}cene \underline{U}nderstanding (CASU)}, illustrated in Figure 1(c), which formalizes auditory scene understanding as a reasoning problem over structured interactions between sounds of speech, events, and background.
To evaluate this capability, we introduce the CASU benchmark, designed to assess whether LALMs can integrate distinct auditory layers to derive contextual meaning at the scene level. CASU employs a scalable semi-synthetic generation pipeline that composes real-world recordings of acoustic scenes (e.g., ambient traffic, room reverberation) and discrete events (e.g., sirens, glass breaking) with semantically controlled synthetic speech, enabling precise control over cross-layer relationships. Building on this pipeline, we define four tasks: contextual reasoning, entity extraction, role inference, and counterfactual reasoning. These tasks explicitly probe models' ability to reason about auditory scenes rather than rely on isolated or incidental cues.

We conduct CASU benchmark on a diverse set of state-of-the-art LALMs. Our results reveal a significant Perception-Understanding Gap: models that achieve near-human performance on single-layer tasks, such as high-fidelity speech transcription, frequently fail on scene understanding tasks that require resolving logical dependencies across other sounds. Furthermore, ablation studies demonstrate that the removal of any single acoustic layer - whether speech, event, or background - leads to a collapse on CASU tasks, confirming that true scene understanding relies on the holistic integration of all auditory dimensions.

In summary, our contributions are threefold:

(1) We introduce Context-Aware Auditory Scene Understanding (CASU), a new paradigm that frames auditory intelligence not as signal filtering, but as understanding and reasoning over the structured interactions among speech, events, and background sounds.

(2) We present a scalable semi-synthetic data generation pipeline that enables precise control over cross-layer contextual relationships, enabling systematical evaluation of scene understanding.

(3) We propose a benchmark with four carefully designed tasks and provide a comprehensive evaluation of state-of-the-art LALMs, revealing persistent limitations in multi-layer auditory understanding and highlighting key directions for future model development.

\begin{figure*}[ht]
    \centering
    \includegraphics[width=\linewidth]{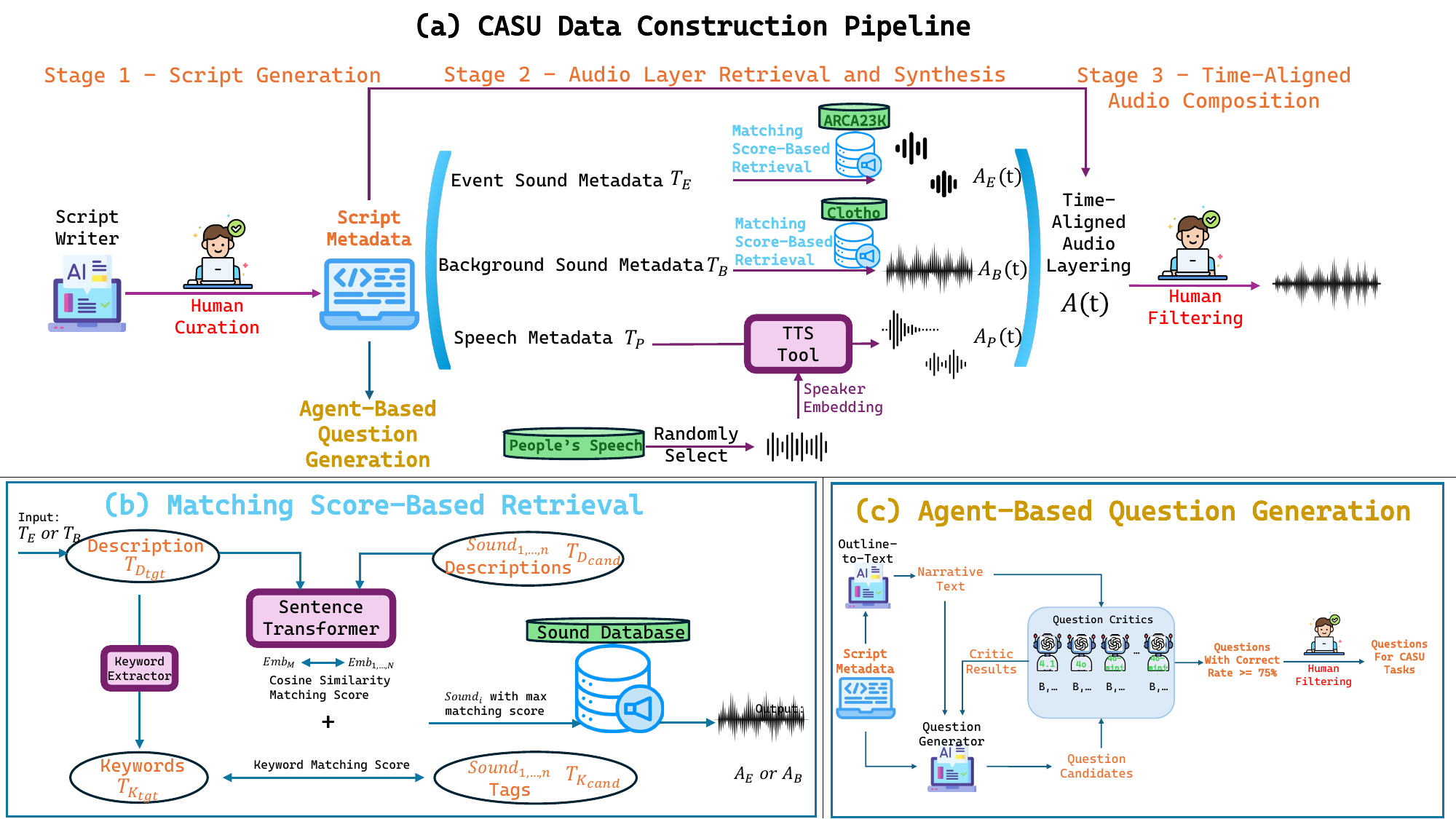} 
    \vspace{-0.7cm}
    \caption{(a) The overview of the CASU data construction pipeline. In the pipeline, we propose (b) Matching-Score Based Retrieval for retriving the most suitable sound based on script, and (c) Agent-Based Question Generation, which can generate questions for all CASU tasks given script metadata.
    }
    \label{fig:pipeline}
\end{figure*}

\begin{figure*}[bt]
    \centering
    \subfloat[\centering \textbf{Contextual Reasoning Task.} Models are required to comprehend the contradiction ($\times$) or confirmation/supplementary ($\checkmark$) between the layers.]{\includegraphics[trim={0.1cm 0.1cm 0.0cm 0.0cm},clip, width=0.97\columnwidth]{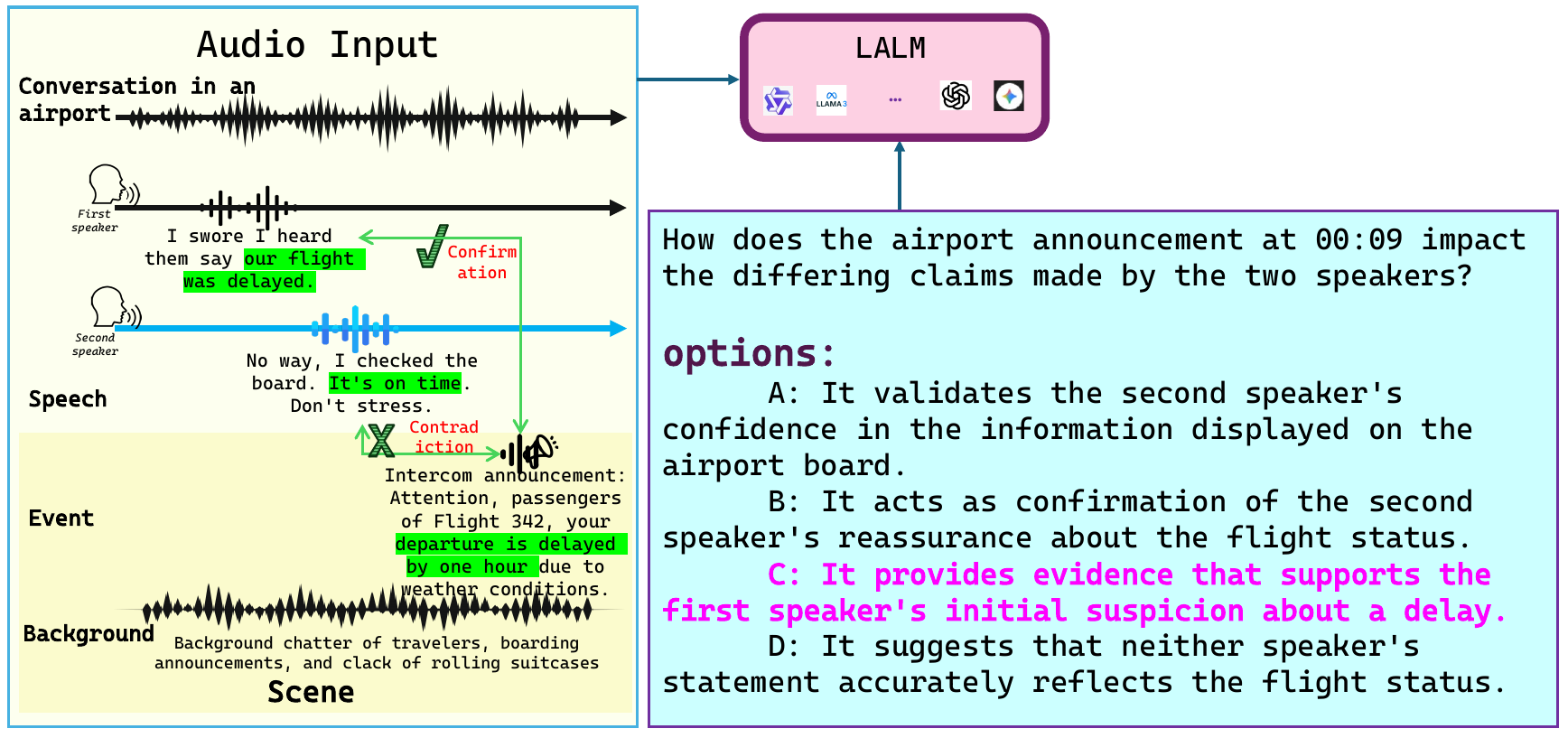} \label{subfig:task1}}%
    \qquad
    \subfloat[\centering \textbf{Entity Extraction Task.} Models are required to determine which entity best explains a particular observation or functional(contextual/causal) relationship within the scene.]{\includegraphics[trim={0.1cm 0.1cm 0.0cm 0.0cm},clip, width=0.97\columnwidth]{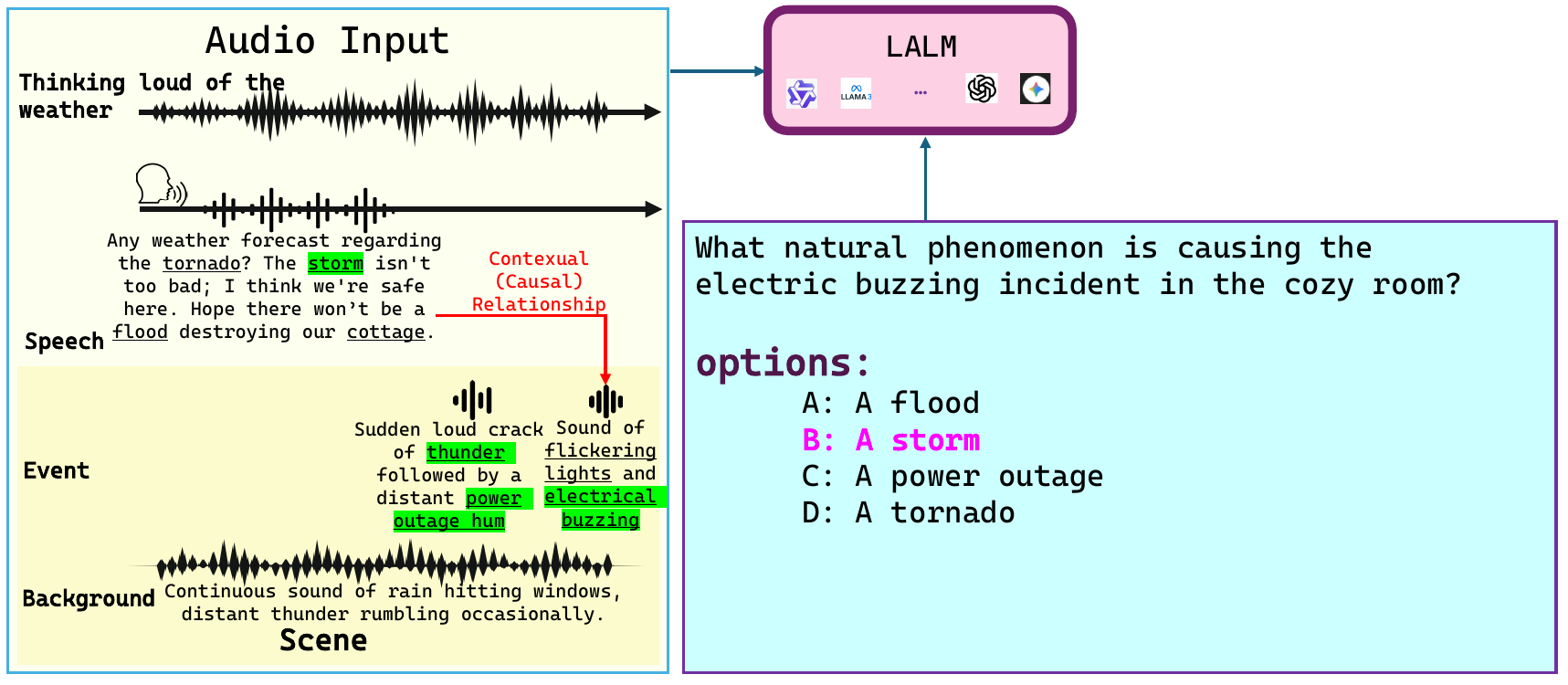}
    \label{subfig:task2}}
      \qquad
        \subfloat[\centering \textbf{Role Inference Task.} Models are required to infer contextual roles or interpersonal relationships from the combination of speech and scene context.]{\includegraphics[trim={0.1cm 0.1cm 0.0cm 0.0cm},clip, width=0.97\columnwidth]{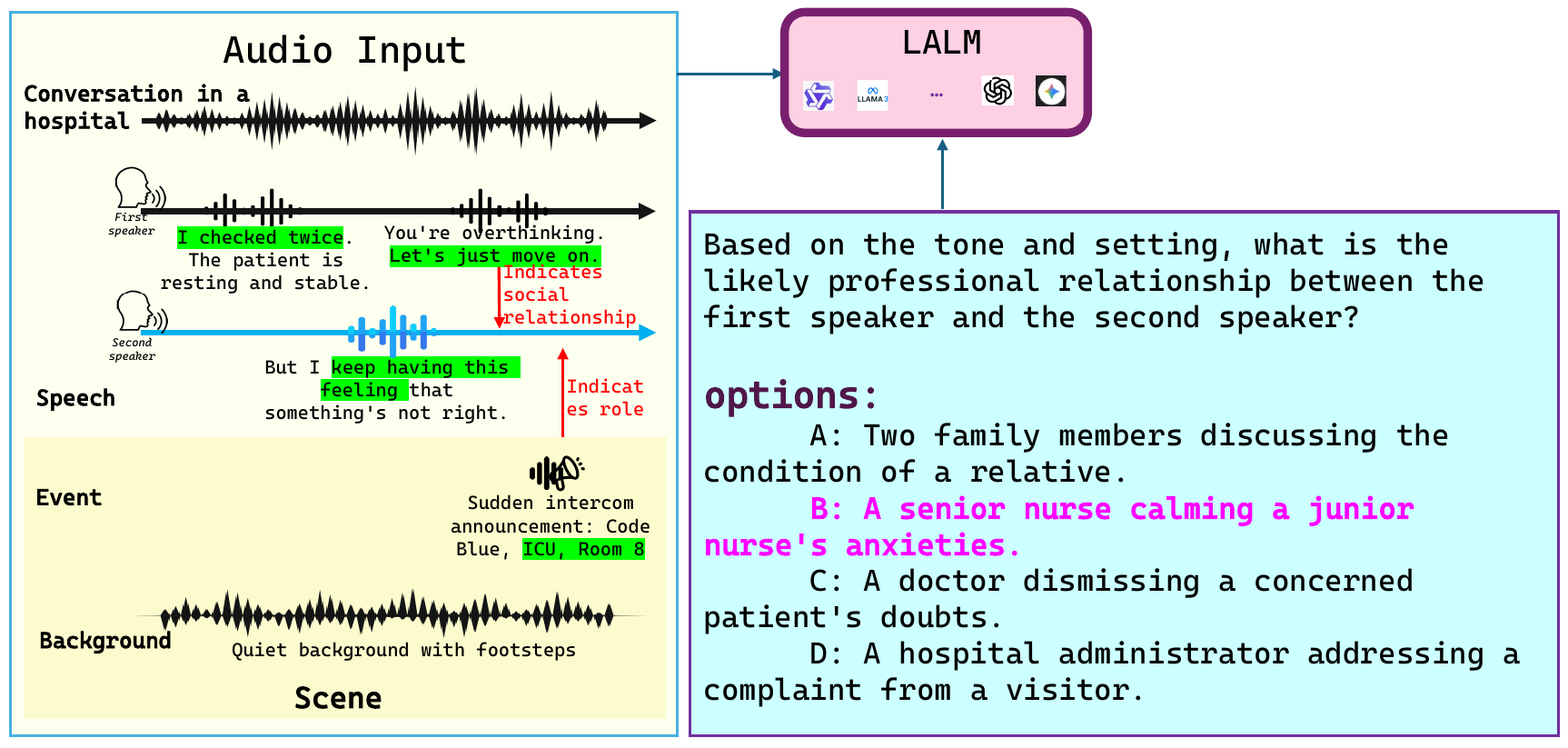}
        \label{subfig:task3}}%
    \qquad
    \subfloat[\centering \textbf{Counterfactual Reasoning Task.} Models are required to reason about causal dependencies within the scene, and how a key interpretation or conclusion would change under an hypothetically altered acoustic evidence.]{\includegraphics[trim={0.1cm 0.1cm 0.0cm 0.0cm},clip, width=0.97\columnwidth]{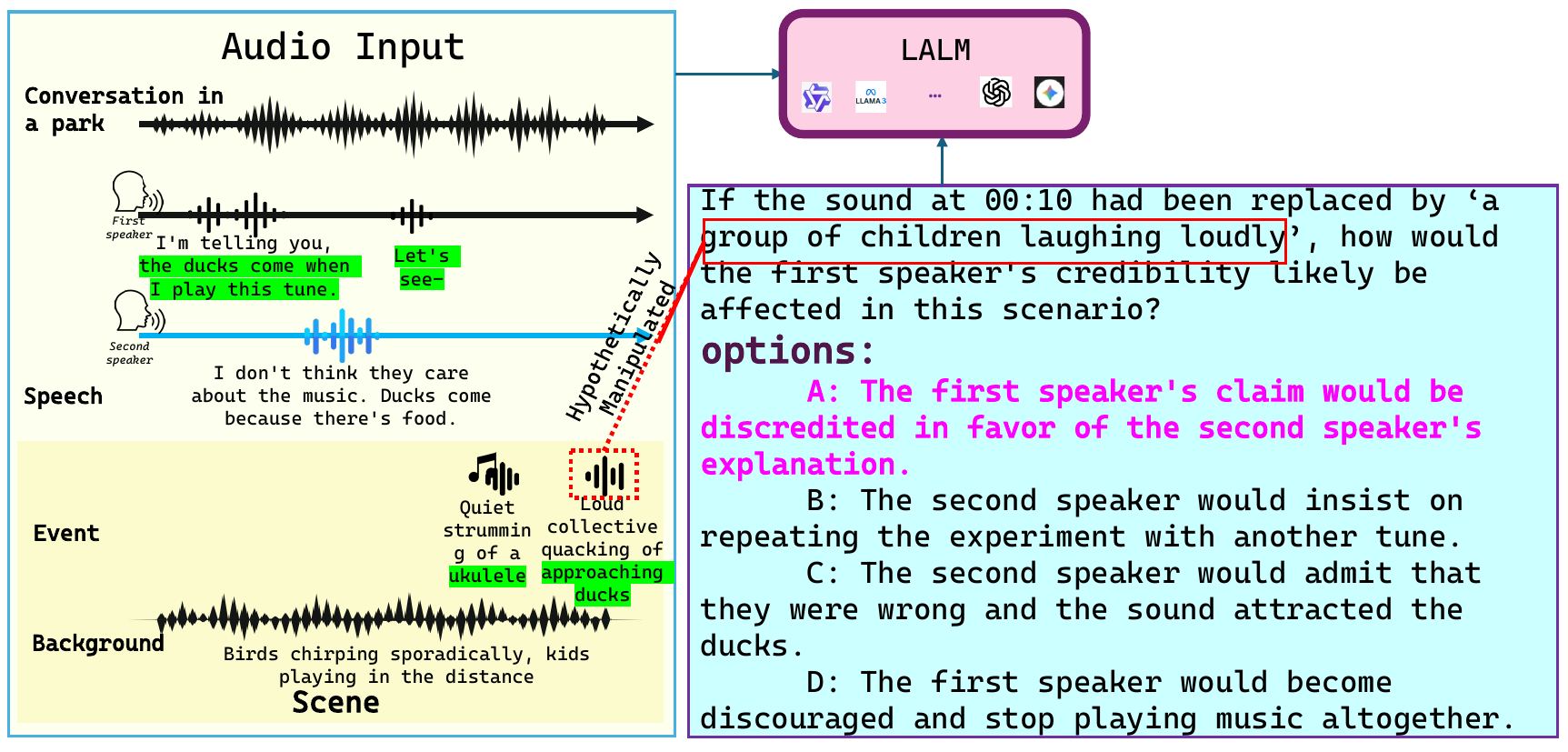}
    \label{subfig:task4}}%
    \vspace{-0.3cm}
    \caption{\textbf{CASU Task Suite.} We introduce four new tasks, Contextual Reasoning, Entity Extraction, Role Inference, and Counterfactual Reasoning, which test different capabilities of LALMs on scene understanding.} %
    \label{fig:casu_tasks}
\end{figure*}

\section{The CASU Benchmark}

\subsection{Dataset Construction}
\label{sec:dataset}

To rigorously evaluate context-aware auditory sense understanding, we require a dataset that offers precise control over the semantic relationships between acoustic layers. Natural 'in-the-wild' audio recordings rarely provide temporal annotations or semantic relationships among speech, events, and background sounds, making them unsuitable for evaluating fine-grained auditory context-aware reasoning. To address this, we introduce a scalable, semi-synthetic data generation pipeline (Fig.~\ref{fig:pipeline}). This pipeline operates in a three-stage procedure: Script Generation, Layer Retrieval \& Synthesis, and Time-Aligned Composition.

\paragraph{Stage 1: Script Generation and Semantic Blueprint.} 
The script of each audio sample is a structured JSON script, generated by a GPT-4o\cite{achiam2023gpt}. We prompt the GPT to synthesize a complex auditory scenario defined by the superposition of three distinct semantic layers:
\begin{enumerate}
    \item \textbf{Background Sound ($T_\mathbf{B}$):} The continuous environmental sound (e.g., \textit{Indoor Private: Dimly Lit Basement. Constant low hum of a generator.}).
    \item \textbf{Speech ($T_\mathbf{P}$):} Human Speech Content.
    \item \textbf{Acoustic Events ($T_\mathbf{E}$):} Discrete, transient sounds that occur in the scene (e.g., \textit{sudden loud bang}).
\end{enumerate}
As shown in the script example (Appendix A Fig~\ref{fig:script_exm}), each component is assigned a text description $T_D$, a start timestamp $t_{\text{start}}$, and an end timestamp $t_{\text{end}}$.
These scripts serve as the semantic blueprint that explicitly states the narrative contextual relationships across layers, which later enable systematic evaluation of scene understanding. A \textbf{Human Curation} step follows generation to filter for logical consistency and narrative reality.

\paragraph{Stage 2: Audio Layer Retrieval and Synthesis.} 
Once the script is finalized, we instantiate the auditory layers using a hybrid approach of Speech Layer Synthesis and Non-Speech Layer Retrieval.

\textit{Speech Layer Synthesis ($\mathbf{A}_{\mathbf{P}}$).} To generate naturalistic speech, we utilize Zonos\footnote{https://github.com/Zyphra/Zonos}, a high-fidelity TTS tools. To ensure speaker diversity across the benchmark, we condition the TTS model on random speaker embeddings sampled from the \textbf{People’s Speech} dataset\cite{galvez2021people}. 

\textit{Non-Speech Layer Retrieval ($\mathbf{A}_{\mathbf{B}}, \mathbf{A}_{\mathbf{E}}$).} For the background sound and event sound, we employ a \textbf{Matching Score-Based Retrieval} system (Fig.~\ref{fig:pipeline}b) to select the optimal audio candidates from source datasets (\textbf{Clotho}\cite{drossos2020clotho} for background scenes and \textbf{ARCA23K}\cite{Iqbal2021} for short-duration events). 

\textbf{Matching Score-Based Retrieval.} (Fig.~\ref{fig:pipeline}b) Given a target description $T_{D_{\text{tgt}}}$ describing the sound $T_B$ or $T_E$, 
we retrieve the most suitable candidate from a sound database where each audio sample is annotated with (i) a text description $T_{D_{\text{cand}}}$ and (ii) a set of categorical tags $T_{K_{\text{cand}}}$ that characterize its acoustic class. We first extract a set of keywords $T_{K_{\text{tgt}}}$ from $T_{D_{\text{tgt}}}$. Then, we compute semantic similarity at both description-level and keyword-level, as categorical tags are particularly informative for defining and disambiguating sound types (e.g., \textit{glass breaking}, \textit{engine}, \textit{laughter}). The matching score $S_{\text{match}}$ is thus formulated as a weighted combination of both similarity scores:
\vspace{-0.2cm}
\begin{equation}
\begin{split}
S_{\text{match}} =
\alpha \cdot \text{Sim}_{\cos}\left( f_{\theta}(T_{D_{\text{tgt}}}), f_{\theta}(T_{D_{\text{cand}}}) \right)
+ \\
(1-\alpha) \cdot \text{Sim}_{\cos}\left( f_{\theta}(T_{K_{\text{tgt}}}), f_{\theta}(T_{K_{\text{cand}}}) \right)
\end{split}
\end{equation}

\vspace{-0.2cm}
where $f_{\theta}(\cdot)$ denotes Sentence Transformer\cite{reimers2019sentence} that maps text to semantic vectors, and $\text{Sim}_{\cos}$ represents the cosine similarity.
The hyperparameter $\alpha$ balances semantic intent with keyword matching. The candidate with the maximal $S_{\text{match}}$ is selected for synthesis, denoted as $\mathbf{A}_{\mathbf{B}}$ or $\mathbf{A}_{\mathbf{E}}$

\paragraph{Stage 3: Time-Aligned Audio Composition.} 
The final audio stream $\mathbf{A}(t)$ is constructed simulating natural recording, i.e, treating the retrieved Background Sound as the base layer and aligning the Speech and Event layers at their designated temporal intervals defined in the script:
\vspace{-0.4cm}
\begin{equation}
\begin{split}
\mathbf{A}(t) = \mathbf{A}_{\mathbf{B}}(t) + \sum_{i=1}^{N_p} \mathbf{A}_{\mathbf{P}}^{(i)}(t-t_{P:\text{start}}^{(i)}) \cdot \mathbb{I}_{[t_{P:\text{start}}^{(i)}, t_{P:\text{end}}^{(i)}]} \\
+ \sum_{j=1}^{N_e} \mathbf{A}_{\mathbf{E}}^{(j)}(t-t_{E:\text{start}}^{(j)}) \cdot \mathbb{I}_{[t_{E:\text{start}}^{(j)}, t_{E:\text{end}}^{(j)}]}
\end{split}
\end{equation}

\vspace{-0.2cm}
where $\mathbb{I}_{[t_{P:\text{start}}, t_{P:\text{end}}]}$ and $\mathbb{I}_{[t_{E:\text{start}}, t_{E:\text{end}}]}$ are indicator functions that are 1 during the active interval of the specific speech segment or event and 0 otherwise. Following composition, a \textbf{Human Filtering} stage is applied to verify that the mixing levels are balanced and that the synthetic composition retains the logical causality intended by the script writer.

\subsection{Tasks and Questions}
\label{sec:task_suite}

A core objective of the CASU benchmark is to evaluate whether LALMs can interpret the contextual logic between speech, background, and events embedded within a complex scene. To this end, besides the \textbf{scene description} task, which requires LALM to recognize and describe all events and transcribe speeches in the audio, we also design four tasks that require models to integrate all contextual acoustic evidence and reason about logical, causal, and social aspects of the audio. These tasks move beyond traditional perceptual benchmarks, challenging models to perform deep semantic inference in auditory contexts where background sounds and events may either constrain or supplement speech topics or assertions.
Fig.~\ref{fig:casu_tasks} provides representative examples for each task. In all cases, LALMs are given an auditory input $\mathbf{A}(t)$ of a scene and a text question $Q$ (Appendix A Table~\ref{tab:qa_exm1}). 

\textbf{Contextual Reasoning.}
The first task, \emph{Contextual Reasoning}, assesses whether a model can reconcile spoken assertions with the broader acoustic context. In many real scenes, the meaning of a dialogue is contingent on background conditions or event that provide supplementary or contradictory evidence. In the example (Fig.~\ref{fig:casu_tasks}a), a speaker's expressed confidence about a situation may be invalidated by an event cue that indicates otherwise, or conversely, may be demonstrated by a situational announcement.

\textbf{Entity Extraction.}
Real-world audio understanding often involves identifying entities that are not directly stated but are implied by the sounds in the scene. The \emph{Entity Extraction} task is designed to evaluate this capability. Given an audio (Fig.~\ref{fig:casu_tasks}b), the query asks the model to determine which entity in the scene best explains a particular observation or functional relationship within the scene.

\textbf{Role Inference.}
Speeches, together with the sounds in the scene, encode social roles, speaker identities, and relational dynamics. To capture this dimension, the \emph{Role Inference} task evaluates whether models can infer contextual roles or interpersonal relationships from the combination of speech and scene context.
This task involves determining the most possible role for a solo speaker in the scene, or the social relationship between two speakers - for example (Fig.~\ref{fig:casu_tasks}c), whether one speaker's choice of words indicate a professional authority over another, and what occupations of them are indicated by the background or event.

\textbf{Counterfactual Reasoning.}
Finally, we introduce \emph{Counterfactual Reasoning} as a means of evaluating a model’s causal understanding of the auditory world. In this task (Fig.~\ref{fig:casu_tasks}d), the associated question asks how a key interpretation or conclusion would change under an hypothetically altered acoustic evidence.
Counterfactual reasoning lies at the heart of human causal inference: to understand not just what is, but what might have been if conditions were different. By introducing hypothetically counterfactual manipulations into the audio input, this task pushes models to reason about causal dependencies between environmental cues and semantic interpretations in a way that goes beyond descriptive understanding.



\textbf{Question Generation.} 
we employ an agent-based question generation framework for all 4 tasks illustrated in Fig.~\ref{fig:pipeline}c\cite{wang2024mmlu}. Starting from the structured script metadata of each audio scene, an LLM first produces a narrative textual description that state the scene containing temporal and causal relations. Based on the description and script, a question generation agent proposes multiple candidate questions aligned with the target task type. 
A panel of independent LLM-based critic agents will evaluate each candidate question by attempting to solve it under the same input conditions, retaining only those questions for which a sufficient consensus on the correct answer is achieved. Human filtering is followed to ensure logic validity, difficulty and category suitable. We defer statistics of dataset to Appendix A.1.


\begin{table*}[t]
\centering
\caption{CASU Benchmark results with different audio understanding models. \textbf{Bold} indicates best performance, \underline{underline} indicates second-best performance. We report \textit{1 - WER} instead of \textit{WER} for the transcription performance, to make sure higher value represents better transcription performance.}
\vspace{-0.4cm}
\resizebox{\linewidth}{!}{
\begin{tabular}{lcccccccccc}

\toprule
\multirow{3.5}{*}{\textbf{Models}} & \multirow{3.5}{*}{\textbf{Size}} & \multicolumn{4}{c}{\textbf{Scene Description - Perception~~$\uparrow$}} & | &  \multicolumn{4}{c}{\textbf{Scene Understanding - CASU (\%)~~$\uparrow$}}  \\ 
\cmidrule{3-11}
&& \textbf{BLEU-4} & \textbf{BertScore} & \textbf{Event} & \textbf{1~-~WER} & | & \textbf{Contextual} & \textbf{Entity} & \textbf{Role} & \textbf{Counterfactual} \\
&& &  & \textbf{Match} & \textbf{(Transcribe)} & | & \textbf{Reasoning} & \textbf{Extraction} & \textbf{Inference} & \textbf{Reasoning} \\
\midrule
Audio Flamingo Chat\cite{kong2024audio}  & 3B& 3.09 & 0.72 & 0.24& 0.92 & | & 31.82 & 29.17& 34.37 & 34.36 \\
Audio Flamingo 2\cite{ghosh2025audio2}  & 3B& 4.54 & 0.75& 0.28& 0.95 & | & 38.60 & 33.24 & 37.05 & 39.45 \\
Audio Flamingo 3\cite{ghosh2025audio3} &7B & 9.20 & 0.78 & 0.46& 0.95 & | & 56.88 & 45.12 & 45.83 & 53.00\\
\midrule
Qwen2-Audio-Instruct\cite{Qwen2-Audio}  & 7B& 1.66 & 0.76 & 0.41& 0.95 & | & 44.02 & 40.81 & 40.00 & 37.81 \\
Qwen2.5-Omni~\cite{Qwen2.5-Omni}  & 7B& 4.48 & 0.73& 0.44& 0.96 & | & 62.06 & 61.10 & 51.37 & 65.38 \\
Qwen3-Omni-30B-A3B-Instruct\cite{Qwen3-Omni} &30B & \textbf{12.80} & 0.81 & 0.65& 0.96 & | & 71.18 & 68.85 & 63.51 & \underline{74.50}\\
\midrule
LTU\cite{gong2023listen}  & 7B& 2.88 & 0.70& 0.15& 0.90 & | & 23.92 & 21.87& 24.18 & 20.99 \\
LTU-AS\cite{gong_ltuas}  & 7B & 2.59 & 0.71& 0.19& 0.92 & | & 20.26 & 19.19& 20.52 & 21.27 \\
LLaMa-Omni\cite{fang2025llamaomni} & 8B& 1.64 & 0.72& 0.20& \textbf{0.98} & | & 53.32 & 38.27 & 49.09 & 57.60 \\
Mistral Voxtral\cite{liu2025voxtral} & 24B& 1.32 & 0.79 & 0.27& 0.97 & | & 52.52 & 41.67 & 51.64 & 55.99 \\
SALMONN\cite{tang2024salmonn}  & 13B& 1.05 & 0.76& 0.36& 0.90 & | & 43.21 & 50.96 & 39.73 & 46.95 \\
\midrule
GPT-4o Audio\cite{achiam2023gpt} &- & \underline{10.53} & \underline{0.82} & \underline{0.72}& \textbf{0.98} & |  & \textbf{74.02} & \underline{70.98} & \underline{68.58} & \textbf{74.96} \\
Gemini 2.0 Flash\cite{comanici2025gemini} &- & 9.19 & \textbf{0.85} & \textbf{0.75}& \textbf{0.98} & | & \underline{73.30} & \textbf{72.77} & \textbf{69.05} & 73.83\\
\midrule
Qwen3-Omni-30B-A3B-Captioner + Qwen3-30B-A3B-Instruct-2507  & -& 3.61 & 0.77 & 0.62& 0.97 & | & 62.08 & 60.40 & 54.28 & 69.63 \\
Qwen3-Omni-30B-A3B-Captioner + GPT-4o  & -& 3.61 & 0.77 & 0.62& 0.97 & | & 63.03 & 59.70& 54.97 & 68.46 \\
GPT-4o-transcribe + Qwen3-30B-A3B-Instruct-2507 &- & 1.08 & 0.74 & -& \textbf{0.98} & | & 57.95 & 41.34 & 51.59 & 58.17 \\
GPT-4o-transcribe + GPT-4o & -& 1.08 & 0.74 & -&\textbf{0.98} & | & 59.77 & 41.75 & 52.46 & 58.10 \\
\bottomrule
\end{tabular}}
\label{tab:exp_mainresult}
\end{table*}

\section{Experiments}

\textbf{Benchmarking Candidates.}
We compare a range of models for audio understanding, including (i) Qwen-series open-source models, with different versions and variants such as audio model Qwen-2-Audio-Instruct\cite{Qwen2-Audio}, and omni-models Qwen-2.5-Omni\cite{Qwen2.5-Omni} and Qwen-3-Omni-Instruct\cite{Qwen3-Omni}; (ii) audio flamingo series models such as Audio-Flamingo-Chat\cite{kong2024audio}, Audio Flamingo 2\cite{ghosh2025audio2}, and Audio Flamingo 3\cite{ghosh2025audio3}; (iii) other open-source LALMs such as Voxtral\cite{liu2025voxtral} and LLaMa-Omni\cite{fang2025llamaomni}, which are mainly trained on speech data; and SALMONN\cite{tang2024salmonn}, LTU\cite{gong2023listen}, LTU-AS\cite{gong_ltuas}, which are trained on diverse sound data; (iv) closed-source LALMs GPT-4o Audio\cite{achiam2023gpt} and Gemini 2.0 Flash\cite{comanici2025gemini}; (iv) cascaded models with Qwen3-Omni-Captioner or GPT-4o-Transcribe as Captioner and GPT-4o or Qwen3-Instruct as Reansoner. Qwen3-Omni-Captioner has the same parameter as Qwen3-Omni-Instructor but can only be used for audio description. 
Reasoner will receive textual audio description as input and answer CASU questions.

\textbf{Evaluation Metrics.} For the \textbf{scene description task}, we have the narrative textual description of each audio (Fig.~\ref{fig:pipeline}c) as reference text. Therefore, we use BLEU-4\cite{papineni2002bleu} and BertScore\cite{reimers2019sentence} compared to the reference text to evaluate the semantic alignment of audio description. To evaluate the perceptual precision on single layers (sound events, background, speeches), we use LLM-as-a-Judge\cite{zheng2023judging} to estimate the percentage of recognized event in the description compared to the audio script, and Word Error Rate (WER)\cite{ali2018word} for speech transcription. For the four new \textbf{CASU scene understanding tasks} we propose, we report the accuracy on the questions. All questions are multiple-choice tests with 4 options.

\section{Results and Discussions}
\subsection{Main Results}

Table \ref{tab:exp_mainresult} reports the performance of a diverse set of LALMs on the CASU benchmark, covering both perceptual (scene description) and scene understanding tasks. Our analysis reveals that strong single-layer perception is necessary but insufficient for scene understanding.

\textbf{Finding 1: The "Perception-Understanding Gap".}
Our results establish a nuanced relationship between perceptual fidelity and scene reasoning. First, the models achieving the highest accuracy on CASU tasks (\textit{GPT-4o Audio, Gemini 2.0 Flash, Qwen3-Omni-Instruct}) also dominate perceptual metrics (BLEU $>9$, Event Match $>0.6$)
However, a significant "Perception-Understanding Gap" emerges when looking at speech-centric models. Models like \textit{Audio Flamingo 3} and \textit{Mistral Voxtral} achieve near-ceiling speech transcription performance ($1-WER \geq 0.95$) yet lag significantly in scene understanding, often trailing state-of-the-art models by $15\%$. This highlights that high-fidelity transcription is insufficient for CASU; the bottleneck for these models is likely their weaker event perception (Event Match $<0.5$), preventing them from utilizing the acoustic scene as a logical constraint. Conversely, lower-fidelity transcription will lead to large performance drop (\textit{SALMONN}).
Finally, the task remains challenging: even the most capable models achieve $70-74\%$ accuracy, suggesting that current LALMs still struggle to fully resolve the complex inter-layer reasoning required by CASU.
This demonstrates that accurate recognition of any single layer (speech or sound) is insufficient; effective scene understanding requires the holistic integration of speech, events, and background.

\textbf{Finding 2: Joint Processing Beats Cascaded Reasoning.}
Omni-modal models (e.g., \textit{Qwen3-Omni}, \textit{GPT-4o Audio}) consistently outperform cascaded pipelines utilizing comparable backbones (\textit{Qwen3-Omni-Captioner + Qwen3-Instruct}), particularly on the Counterfactual and Contextual Reasoning tasks. Cascaded systems, which convert audio to a textual description before reasoning, suffer from an "information loss" due to insufficient event match ($event match=0.62$) or transcription ($1-WER=0.97$). They inevitably discard subtle acoustic cues, such as the reverberation suggesting a large hall, which are tough to be written down for the audio description task. These lost signals often serve as the pivotal context needed to understand the scene or resolve logical ambiguities in CASU.

\textbf{Finding 3: Different Layers in the Scene Play Different Roles, Not Just Noise.} The performance variance on Entity Extraction vs. Role Inference reveals that CASU requires models to dynamically weigh contributions across auditory layers. Entity Extraction relies more on non-speech acoustic evidence. Models with strong sound recognition (\textit{Qwen2.5-Omni}, \textit{SALMONN}) excel here (\textit{SALMONN} outperforms \textit{Voxtral} by $9\%$ on entity extraction, despite much worse performance on the other tasks), provided speech quality is not severely degraded. Role Inference is linguistically heavy but contextually grounded; it requires the model to infer social dynamics (e.g., "Customer" vs. "Server") which are often signaled by the timing of background events (e.g., a register beep after a sentence). Models with better transcription performance (\textit{Voxtral}, \textit{LLaMa-Omni}) perform good on this task. This validates our core hypothesis: effective LALMs must actively discern when an acoustic layer serves as a semantic anchor for the scene.
Contextual Reasoning and Counterfactual Reasoning are the most demanding, requiring the model to disentangle layers. Models with low event perception performance or transcription performance (\textit{LTU}, \textit{LTU-AS}) collapse on both tasks.


\subsection{Ablation Study on Audio Layers}

To strictly verify the contribution of distinct auditory layers to scene understanding, we conducted a component-wise ablation study using \textit{Qwen2.5-Omni}. We masked the speech ($A_P$), events ($A_E$), and background ($A_B$) layers respectively with Gaussian noise, evaluating performance across all permutation subsets. Table \ref{tab:exp_abs_layer} details the results.

\begin{table}[h]
\centering
\caption{Results of Qwen2.5-Omni on audios without sounds of background($A_B$)/event($A_E$)/speech($A_P$).}
    \vspace{-0.4cm}
\resizebox{\linewidth}{!}{
\begin{tabular}{lcccc}

\toprule
\multirow{2.5}{*}{\textbf{Audio Layers}} & \multicolumn{4}{c}{\textbf{Scene Understanding (\%)~~$\uparrow$}}  \\ 
\cmidrule{2-5}
& \textbf{Contextual} & \textbf{Entity} & \textbf{Role} & \textbf{Counterfactual} \\
$A_P,A_E,A_B$ & \textbf{Reasoning} & \textbf{Extraction} & \textbf{Inference} & \textbf{Reasoning} \\
\midrule
$\times,~~\times,~~\times$ & 20.44 & 17.47 & 22.88 & 27.90 \\
$\times,~~\times,~~\checkmark$ & 24.89 & 26.17 & 31.93 & 21.16 \\
$\times,~~\checkmark,~~\times$ & 26.19 & 37.19 & 31.95 & 23.24 \\
$\times,~~\checkmark,~~\checkmark$ & 26.10 & 35.79 & 33.56 & 20.84 \\
$\checkmark,~~\times,~~\times$ & 57.53 & 53.41 & 49.81 & 55.18 \\
$\checkmark,~~\times,~~\checkmark$ & 58.97 & 54.11 & 49.06 & 56.75 \\
$\checkmark,~~\checkmark,~~\times$ & 60.58 & 60.46 & 51.47 & 64.33 \\
$\checkmark,~~\checkmark,~~\checkmark$ & 62.06 & 61.10 & 51.37 & 65.38 \\
\bottomrule
\end{tabular}}
\label{tab:exp_abs_layer}
\end{table}

\textbf{Finding 4: Speech serves as the semantic foundation, but the Scene provides the logical constraints.} As shown in Table \ref{tab:exp_abs_layer}, speech ($A_P$) is the dominant carrier of information; removing it (rows 1-4) causes a performance collapse across all tasks, with accuracy down to near random guess levels (20-30\%). However, relying on speech alone (Row 5: $A_P, \times, \times$) is insufficient for CASU. The complete removal of the acoustic scene ($A_E$ and $A_B$) results in a significant performance degradation compared to the full model (Row 8), particularly in Counterfactual Reasoning (-10.2\%) and \textit{Entity Extraction} (-7.7\%). This empirically proves that current LALMs do not solve these tasks by merely "guessing" from text; they actively utilize non-speech acoustic cues to ground their reasoning. 

\textbf{Events contribute to entity extraction.} Adding discrete events ($A_E$) to the speech baseline (Row 5 vs. Row 7) yields the largest gain for Entity Extraction ($53.41\% \rightarrow 60.46\%$). This confirms that models locate physical entities primarily through their acoustic signatures (e.g., a "dog" is confirmed by barking, not just the speaker mentioning a pet)

\subsection{Scene Difficulty on Speech}
\begin{figure}[htbp]
  \centering
  \begin{subfigure}{.23\textwidth}
    \centering
    \includegraphics[width=\linewidth]{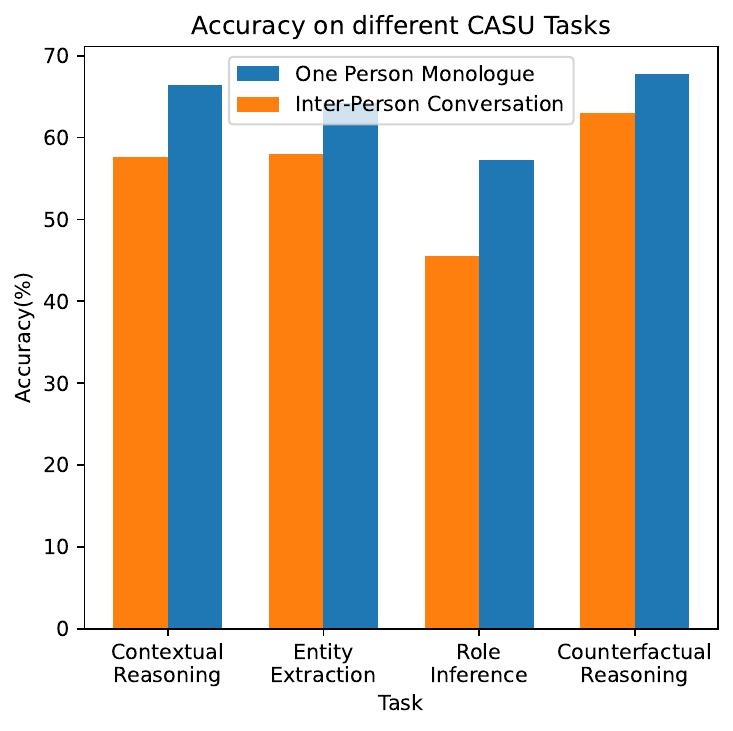}
    \label{fig:speech1}
  \end{subfigure}
  \hfill
  \begin{subfigure}{.23\textwidth}
    \centering
    \includegraphics[width=\linewidth]{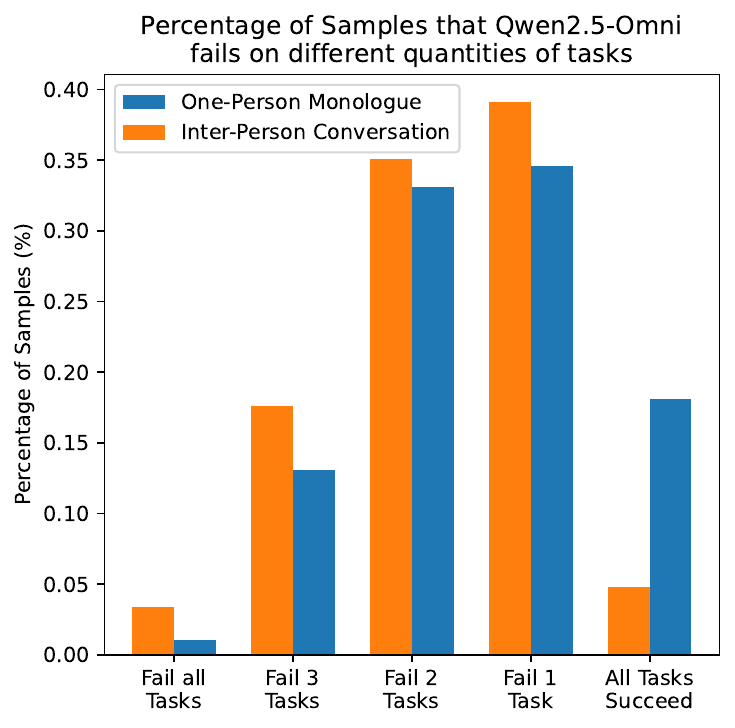}
    \label{fig:speech2}
  \end{subfigure}
  \vspace{-1cm}
  \caption{Results of Qwen2.5-Omni on CASU Tasks on different speech settings for the scene.}
  \label{fig:speechsetting}
\end{figure}

Our dataset contains both two types of speech: One-Person Monologue and Inter-Person Conversation. Fig.~\ref{fig:speechsetting} shows how different speech difficulty influence scene understanding.

\textbf{Finding 5: Multi-Person Scene Setting Increases Understanding Complexity.} We observe a consistent performance degradation when shifting from single-person monologues (Blue bars) to inter-person conversations (Orange bars) across all four tasks. This indicates that conversation scene introduce significant difficulty on integrating other sound in the scene. As in Fig.~\ref{fig:speechsetting}b, models often fail on more percentage of audio samples.
Role Inference task suffers the most, where accuracy drops by approximately $12\%$. This sharp decline validates the hypothesis that social reasoning is harder in conversations than monologue. 

\subsection{Influence of Scene Hints}

To further investigate whether models struggle with percepting the scene, we evaluate performance when an explicit textual hint for the scene place is provided (e.g., "This audio is recorded in a [Location]"). Table \ref{tab:abs_hint} compares the performance with and without this auxiliary information. 
\begin{table}[h]
\centering
\caption{CASU Benchmark results with scene text hint.}
    \vspace{-0.4cm}
\resizebox{\linewidth}{!}{
\begin{tabular}{lccccc}

\toprule
\multirow{2}{*}{\textbf{Model}}&\multirow{2}{*}{\textbf{Prompt}} & \multicolumn{4}{c}{\textbf{Scene Understanding (\%)~~$\uparrow$}}  \\ 
\cmidrule{3-6}
&\textbf{Condition} & \textbf{Contextual} & \textbf{Entity} & \textbf{Role} & \textbf{Counterfactual} \\
&& \textbf{Reasoning} & \textbf{Extraction} & \textbf{Inference} & \textbf{Reasoning} \\
\midrule
\multirow{2}{*}{\textbf{Qwen2.5-Omni}}& with hint & 64.14 & 63.90 & 55.19 & 65.91 \\
& without hint & 62.06 & 61.10 & 51.37 & 65.38 \\
\midrule
\multirow{2}{*}{\textbf{GPT-4o Audio}}& with hint & 74.15 & 72.36 & 70.01 & 74.42 \\
& without hint & 74.02 & 70.98 & 68.58 & 74.96 \\
\bottomrule
\end{tabular}}
\label{tab:abs_hint}
\end{table}

\begin{table*}[htbp]
\centering
\footnotesize
\caption{Illustration of different error types on the CASU scene understanding questions.}
\vspace{-0.4cm}
\resizebox{\linewidth}{!}{
\begin{tabular}{|p{1.5cm}|p{6.7cm}|p{5.8cm}|}
\hline
\textbf{Error Type} & \textbf{Perception or Reasoning Error} & \textbf{Self-Reasoning Error} \\

\hline
\textbf{Question}
& 
What is the most plausible relationship between the first and the second speaker?
\newline A: Two strangers who happened to walk the same forest trail by accident.
\newline B: A parent calming their child's nighttime fears during a camping trip.
\newline C: Two friends on an adventurous night hike, one reassuring the other. \newline D: A wildlife expert guiding a nervous tourist on a nature walk.
& 
Which particular sound causes the interruption of the tranquil cooking environment in the kitchen?  
\newline A: The sound of alarm,    
\newline B: The blender malfunction noise,
\newline C: The beep from the smoke detector,    
\newline D: The beep from the kettle with boiling water
\\
\hline
\textbf{Wrong Answer}
& 
Answer: B. Justification: The audio contains a \underline{child's voice expressing fear} \textcolor{mygreen}{(\textit{Perception Error})}, which is a \underline{key indicator} \textcolor{mygreen}{(\textit{No Causal Relationship})} of a parent-child relationship.
& 
Answer: C. Justification: Across all options, the beep from the smoke detector is more likely to interrupt ... This is evident from \underline{the sudden, loud, and disruptive nature}  \textcolor{mygreen}{(\textit{Not cue in audio})} of the smoke detector's beep
\\
\hline
\textbf{Correct Answer}
& 
Answer: C. Justification: ...the interaction is 'alarm/dismissive' but the \textbf{conversation is comforting, suggesting familiarity and comfort (Friends)}. ...No adult-like dialogue so B is incorrect...
& 
Answer: D. Justification: ... This is \textbf{evident from the sound of boiling water in the audio}, ..., causing a disruption in the otherwise peaceful scene.
\\
\hline
\end{tabular}
}
\label{tab:error_type}
\vspace{-0.3cm}
\end{table*}

\textbf{Finding 6: Textual Hint Reduces Acoustic Ambiguity.} Providing an informative scene hint consistently improves performance, particularly for open-source models. \textit{Qwen2.5-Omni} sees a performance boost across all tasks, with the most gain in Role Inference (+3.82\%). 
This suggests that once the scene is explicitly defined (e.g., "hospital"), the model can more effectively retrieve the associated social schemas (e.g., Doctor/Patient dynamics) for reasoning. However, the benefit of textual hints is less for \textit{GPT-4o Audio}. This indicates that stronger LALMs likely possess a robust internal representation of the auditory scene; they may have already implicitly inferred the context from the audio alone, rendering the textual hint redundant. 



\subsection{Error Analysis}

We present two different types of errors in Table~\ref{tab:error_type}. The first one is due to LALMs' perception or reasoning error, and the second type is because the LALMs may use the information in the question or the internal knowledge for reasoning, not the cues or information in the audio.

\section{Related Work}
\label{sec:related_work}

\textbf{Large Audio-Language Models (LALMs).} Recent advancements in LALMs have shifted from specialized single-task models to unified "omni-modal" architectures capable of processing speech, music, and sound events\cite{Qwen3-Omni}. Early works like SALMONN\cite{tang2024salmonn} and Qwen-Audio\cite{Qwen2-Audio} demonstrated strong capabilities in separate perception tasks, such as automatic speech recognition (ASR) and audio captioning. More recently, models like GPT-4o Audio\cite{achiam2023gpt}, Gemini 2.0 Flash\cite{comanici2025gemini}, and Audio Flamingo\cite{ghosh2025audio3} have pushed the boundaries of general auditory reasoning. Despite these strides, most existing LALMs are trained and evaluated primarily on loosely coupled audio streams. They often treat non-speech sounds as secondary background signals rather than semantic anchors, limiting their ability to resolve scenarios where the acoustic scene (e.g., an emergency siren) fundamentally alters the logical interpretation of the spoken content.

\noindent \textbf{Audio Understanding and Reasoning Benchmarks.}
Evaluating holistic audio intelligence remains a significant challenge. Initial benchmarks focused on specific verticals: OpenASQA for speech \cite{gong_ltuas}, CompA\cite{ghosh2023compa} for sound events, and MusicBench\cite{melechovsky2024mustango} for musical understanding. Recognizing the need for broader evaluation, recent suites like AIR-Bench\cite{yang2024air} and AudioBench\cite{wang2025audiobench} have unified diverse datasets to test fundamental tasks like ASR and scene classification. To probe deeper reasoning, MMAU\cite{sakshi2024mmau} introduced massive multi-disciplinary QA pairs across speech, sound, and music, while MMAU-Pro\cite{kumar2025mmau} and SAKURA\cite{yang2025sakura} tested hierarchical reasoning. Similarly, Dynamic-SUPERB\cite{huang2024dynamic} expanded instruction-tuned evaluation to cover over 180 tasks. While these benchmarks assess the breadth of audio knowledge, they often treat mixed audio as a loose collection of signals. They rarely evaluate Context-Aware case, where the model must actively discern whether the background acts as noise or as a critical semantic anchor that alters the logic of the foreground speech. Our work fills this gap by formalizing scene-level interaction and reasoning.

\section{Conclusion}
We introduced Context-Aware Auditory Scene Understanding (CASU), a new paradigm that shifts audio evaluation from isolated perception to holistic reasoning over interacting acoustic layers. Our findings demonstrate that true auditory intelligence requires models to actively discern the semantic utility of non-speech sounds rather than treating them as mere noise. We hope CASU serves as a critical stepping stone for future LALMs to bridge the divide between recognizing signals and understanding real-world auditory scenes.

\section{Limitations}
Despite the rigorous semi-synthetic data generation pipeline together with human-in-the-lop filtering, our CASU benchmark might remain synthetic compared to fully human-labeled and curated datasets such as MMAR \cite{ma2025mmar}, MMAU-PRO \cite{kumar2025mmau}. However, our underlying pipeline is modular and configurable by design, facilitating future adaptation and generalization towards different domains and settings.

Our provided dataset contains audio scenes under 30 seconds, due to current LALM's limitation on processing audios. Some scene might be "expedited" (i.e, events happen soon after the speech, but in real world there might be a time gap), as we are trying to crowd real-world scene into 30 seconds clip. Without loss of generality, our proposed pipeline can simulate real world scene. And our pipeline can definitely generate longer audios if furture LALMs can process longer audios.

In addition, due to LALMs on audio processing limit ($<30s$), introducing conversations with more than 2 persons make the audio hard-to-understand for human (lots of speech overlapping). Therefore, current audios are confined to 1-person monologue and 2-person conversational scenarios. However, our pipeline can serve as a robust baseline for future complex interactions. The constraint of 2-person conversation facilitates an incremental progression towards natural human interactions without introducing unnecessary confounding variables associated with intricate multi-party conversational settings. We leave the expansion to support N-person conversational dynamics for future works.

\bibliography{custom}

@inproceedings{ghosh2025audio3,
  title={Audio Flamingo 3: Advancing Audio Intelligence with Fully Open Large Audio Language Models},
  author={Ghosh, Sreyan and Goel, Arushi and Kim, Jaehyeon and Kumar, Sonal and Kong, Zhifeng and Lee, Sang-gil and Yang, Chao-Han Huck and Duraiswami, Ramani and Manocha, Dinesh and Valle, Rafael and Catanzaro, Bryan},
  booktitle={The Thirty-ninth Annual Conference on Neural Information Processing Systems},
  year={2025},
  url={https://openreview.net/forum?id=FjByDpDVIO}
}

@inproceedings{kong2024audio,
  title={Audio Flamingo: A Novel Audio Language Model with Few-Shot Learning and Dialogue Abilities},
  author={Kong, Zhifeng and Goel, Arushi and Badlani, Rohan and Ping, Wei and Valle, Rafael and Catanzaro, Bryan},
  booktitle={International Conference on Machine Learning},
  pages={25125--25148},
  year={2024},
  organization={PMLR}
}

@inproceedings{ghosh2025audio2,
  title={Audio Flamingo 2: An Audio-Language Model with Long-Audio Understanding and Expert Reasoning Abilities},
  author={Ghosh, Sreyan and Kong, Zhifeng and Kumar, Sonal and Sakshi, S and Kim, Jaehyeon and Ping, Wei and Valle, Rafael and Manocha, Dinesh and Catanzaro, Bryan},
  booktitle={Forty-second International Conference on Machine Learning},
  year={2025},
  url={https://openreview.net/forum?id=xWu5qpDK6U}
}

@inproceedings{
  fang2025llamaomni,
  title={{LL}a{MA}-{O}mni: Seamless Speech Interaction with Large Language Models},
  author={Qingkai Fang and Shoutao Guo and Yan Zhou and Zhengrui Ma and Shaolei Zhang and Yang Feng},
  booktitle={The Thirteenth International Conference on Learning Representations},
  year={2025},
  url={https://openreview.net/forum?id=PYmrUQmMEw}
}

@article{Qwen3-Omni,
  title={Qwen3-Omni Technical Report},
  author={Jin Xu and Junyang Lin},
  journal={arXiv preprint arXiv:2509.17765},
  year={2025}
}

@article{Qwen2.5-Omni,
  title={Qwen2.5-Omni Technical Report},
  author={Jin Xu},
  journal={arXiv preprint arXiv:2503.20215},
  year={2025}
}

@inproceedings{fuentes2022urban,
  title={Urban sound \& sight: Dataset and benchmark for audio-visual urban scene understanding},
  author={Fuentes, Magdalena and Steers, Bea and Zinemanas, Pablo and Rocamora, Martin and Bondi, Luca and Wilkins, Julia and Shi, Qianyi and Hou, Yao and Das, Samarjit and Serra, Xavier and others},
  booktitle={ICASSP 2022-2022 IEEE International Conference on Acoustics, Speech and Signal Processing (ICASSP)},
  pages={141--145},
  year={2022},
  organization={IEEE}
}

@inproceedings{huang2024dynamic,
  title={Dynamic-superb: Towards a dynamic, collaborative, and comprehensive instruction-tuning benchmark for speech},
  author={Huang, Chien-yu and Lu, Ke-Han and Wang, Shih-Heng and Hsiao, Chi-Yuan and Kuan, Chun-Yi and Wu, Haibin and Arora, Siddhant and Chang, Kai-Wei and Shi, Jiatong and Peng, Yifan and others},
  booktitle={ICASSP 2024-2024 IEEE International Conference on Acoustics, Speech and Signal Processing (ICASSP)},
  pages={12136--12140},
  year={2024},
  organization={IEEE}
}

@inproceedings{wang2025audiobench,
  title={Audiobench: A universal benchmark for audio large language models},
  author={Wang, Bin and Zou, Xunlong and Lin, Geyu and Sun, Shuo and Liu, Zhuohan and Zhang, Wenyu and Liu, Zhengyuan and Aw, AiTi and Chen, Nancy},
  booktitle={Proceedings of the 2025 Conference of the Nations of the Americas Chapter of the Association for Computational Linguistics: Human Language Technologies (Volume 1: Long Papers)},
  pages={4297--4316},
  year={2025}
}

@article{yang2024air,
  title={Air-bench: Benchmarking large audio-language models via generative comprehension},
  author={Yang, Qian and Xu, Jin and Liu, Wenrui and Chu, Yunfei and Jiang, Ziyue and Zhou, Xiaohuan and Leng, Yichong and Lv, Yuanjun and Zhao, Zhou and Zhou, Chang and others},
  journal={arXiv preprint arXiv:2402.07729},
  year={2024}
}

@inproceedings{melechovsky2024mustango,
  title={Mustango: Toward controllable text-to-music generation},
  author={Melechovsky, Jan and Guo, Zixun and Ghosal, Deepanway and Majumder, Navonil and Herremans, Dorien and Poria, Soujanya},
  booktitle={Proceedings of the 2024 Conference of the North American Chapter of the Association for Computational Linguistics: Human Language Technologies (Volume 1: Long Papers)},
  pages={8293--8316},
  year={2024}
}

@article{ghosh2023compa,
  title={Compa: Addressing the gap in compositional reasoning in audio-language models},
  author={Ghosh, Sreyan and Seth, Ashish and Kumar, Sonal and Tyagi, Utkarsh and Evuru, Chandra Kiran and Ramaneswaran, S and Sakshi, S and Nieto, Oriol and Duraiswami, Ramani and Manocha, Dinesh},
  journal={arXiv preprint arXiv:2310.08753},
  year={2023}
}

@article{lu2015context,
  title={Context-based environmental audio event recognition for scene understanding},
  author={Lu, Tong and Wang, Gongyou and Su, Feng},
  journal={Multimedia Systems},
  volume={21},
  number={5},
  pages={507--524},
  year={2015},
  publisher={Springer}
}

@inproceedings{du2025crab,
  title={Crab: A unified audio-visual scene understanding model with explicit cooperation},
  author={Du, Henghui and Li, Guangyao and Zhou, Chang and Zhang, Chunjie and Zhao, Alan and Hu, Di},
  booktitle={Proceedings of the Computer Vision and Pattern Recognition Conference},
  pages={18804--18814},
  year={2025}
}

@article{Qwen2-Audio,
  title={Qwen2-Audio Technical Report},
  author={Chu, Yunfei and Xu, Jin and Yang, Qian and Wei, Haojie and Wei, Xipin and Guo,  Zhifang and Leng, Yichong and Lv, Yuanjun and He, Jinzheng and Lin, Junyang and Zhou, Chang and Zhou, Jingren},
  journal={arXiv preprint arXiv:2407.10759},
  year={2024}
}

@inproceedings{
  tang2024salmonn,
  title={SALMONN: Towards Generic Hearing Abilities for Large Language Models},
  author={Changli Tang and Wenyi Yu and Guangzhi Sun and Xianzhao Chen and Tian Tan and Wei Li and Lu Lu and Zejun MA and Chao Zhang},
  booktitle={The Twelfth International Conference on Learning Representations},
  year={2024},
  url={https://openreview.net/forum?id=14rn7HpKVk}
}

@article{liu2025voxtral,
  title={Voxtral},
  author={Liu, Alexander H and Ehrenberg, Andy and Lo, Andy and Denoix, Cl{\'e}ment and Barreau, Corentin and Lample, Guillaume and Delignon, Jean-Malo and Chandu, Khyathi Raghavi and von Platen, Patrick and Muddireddy, Pavankumar Reddy and others},
  journal={arXiv preprint arXiv:2507.13264},
  year={2025}
}

@inproceedings{gong_ltuas,
  title={Joint Audio and Speech Understanding},
  author={Gong, Yuan and Liu, Alexander H and Luo, Hongyin and Karlinsky, Leonid and Glass, James},
  year={2023},
  booktitle={2023 IEEE Automatic Speech Recognition and Understanding Workshop (ASRU)},
}

@article{gong2023listen,
  title={Listen, Think, and Understand},
  author={Gong, Yuan and Luo, Hongyin and Liu, Alexander H and Karlinsky, Leonid and Glass, James},
  journal={arXiv preprint arXiv:2305.10790},
  year={2023}
}

@article{achiam2023gpt,
  title={Gpt-4 technical report},
  author={Achiam, Josh and Adler, Steven and Agarwal, Sandhini and Ahmad, Lama and Akkaya, Ilge and Aleman, Florencia Leoni and Almeida, Diogo and Altenschmidt, Janko and Altman, Sam and Anadkat, Shyamal and others},
  journal={arXiv preprint arXiv:2303.08774},
  year={2023}
}

@article{comanici2025gemini,
  title={Gemini 2.5: Pushing the frontier with advanced reasoning, multimodality, long context, and next generation agentic capabilities},
  author={Comanici, Gheorghe and Bieber, Eric and Schaekermann, Mike and Pasupat, Ice and Sachdeva, Noveen and Dhillon, Inderjit and Blistein, Marcel and Ram, Ori and Zhang, Dan and Rosen, Evan and others},
  journal={arXiv preprint arXiv:2507.06261},
  year={2025}
}

@article{sakshi2024mmau,
  title={Mmau: A massive multi-task audio understanding and reasoning benchmark},
  author={Sakshi, S and Tyagi, Utkarsh and Kumar, Sonal and Seth, Ashish and Selvakumar, Ramaneswaran and Nieto, Oriol and Duraiswami, Ramani and Ghosh, Sreyan and Manocha, Dinesh},
  journal={arXiv preprint arXiv:2410.19168},
  year={2024}
}

@article{ma2025mmar,
  title={MMAR: A Challenging Benchmark for Deep Reasoning in Speech, Audio, Music, and Their Mix},
  author={Ma, Ziyang and Ma, Yinghao and Zhu, Yanqiao and Yang, Chen and Chao, Yi-Wen and Xu, Ruiyang and Chen, Wenxi and Chen, Yuanzhe and Chen, Zhuo and Cong, Jian and others},
  journal={arXiv preprint arXiv:2505.13032},
  year={2025}
}

@article{kumar2025mmau,
  title={Mmau-pro: A challenging and comprehensive benchmark for holistic evaluation of audio general intelligence},
  author={Kumar, Sonal and Sedl{\'a}{\v{c}}ek, {\v{S}}imon and Lokegaonkar, Vaibhavi and L{\'o}pez, Fernando and Yu, Wenyi and Anand, Nishit and Ryu, Hyeonggon and Chen, Lichang and Pli{\v{c}}ka, Maxim and Hlav{\'a}{\v{c}}ek, Miroslav and others},
  journal={arXiv preprint arXiv:2508.13992},
  year={2025}
}

@article{yang2025sakura,
  title={Sakura: On the multi-hop reasoning of large audio-language models based on speech and audio information},
  author={Yang, Chih-Kai and Ho, Neo and Piao, Yen-Ting and Lee, Hung-yi},
  journal={arXiv preprint arXiv:2505.13237},
  year={2025}
}

@inproceedings{Iqbal2021,
    author = {Iqbal, T. and Cao, Y. and Bailey, A. and Plumbley, M. D. and Wang, W.},
    title = {{ARCA23K}: An audio dataset for investigating open-set label noise},
    booktitle = {Proceedings of the Detection and Classification of Acoustic Scenes and Events 2021 Workshop (DCASE2021)},
    pages = {201--205},
    year = {2021},
    address = {Barcelona, Spain},
}

@article{wang2024mmlu,
  title={Mmlu-pro: A more robust and challenging multi-task language understanding benchmark},
  author={Wang, Yubo and Ma, Xueguang and Zhang, Ge and Ni, Yuansheng and Chandra, Abhranil and Guo, Shiguang and Ren, Weiming and Arulraj, Aaran and He, Xuan and Jiang, Ziyan and others},
  journal={Advances in Neural Information Processing Systems},
  volume={37},
  pages={95266--95290},
  year={2024}
}

@inproceedings{drossos2020clotho,
  title={Clotho: An audio captioning dataset},
  author={Drossos, Konstantinos and Lipping, Samuel and Virtanen, Tuomas},
  booktitle={ICASSP 2020-2020 IEEE International Conference on Acoustics, Speech and Signal Processing (ICASSP)},
  pages={736--740},
  year={2020},
  organization={IEEE}
}

@article{galvez2021people,
  title={The people's speech: A large-scale diverse english speech recognition dataset for commercial usage},
  author={Galvez, Daniel and Diamos, Greg and Ciro, Juan and Cer{\'o}n, Juan Felipe and Achorn, Keith and Gopi, Anjali and Kanter, David and Lam, Maximilian and Mazumder, Mark and Reddi, Vijay Janapa},
  journal={arXiv preprint arXiv:2111.09344},
  year={2021}
}

@article{reimers2019sentence,
  title={Sentence-bert: Sentence embeddings using siamese bert-networks},
  author={Reimers, Nils and Gurevych, Iryna},
  journal={arXiv preprint arXiv:1908.10084},
  year={2019}
}

@inproceedings{papineni2002bleu,
  title={Bleu: a method for automatic evaluation of machine translation},
  author={Papineni, Kishore and Roukos, Salim and Ward, Todd and Zhu, Wei-Jing},
  booktitle={Proceedings of the 40th annual meeting of the Association for Computational Linguistics},
  pages={311--318},
  year={2002}
}

@article{zheng2023judging,
  title={Judging llm-as-a-judge with mt-bench and chatbot arena},
  author={Zheng, Lianmin and Chiang, Wei-Lin and Sheng, Ying and Zhuang, Siyuan and Wu, Zhanghao and Zhuang, Yonghao and Lin, Zi and Li, Zhuohan and Li, Dacheng and Xing, Eric and others},
  journal={Advances in neural information processing systems},
  volume={36},
  pages={46595--46623},
  year={2023}
}

@inproceedings{ali2018word,
  title={Word error rate estimation for speech recognition: e-WER},
  author={Ali, Ahmed and Renals, Steve},
  booktitle={Proceedings of the 56th Annual Meeting of the Association for Computational Linguistics (Volume 2: Short Papers)},
  pages={20--24},
  year={2018},
  organization={Association for Computational Linguistics (ACL)}
}

@article{clark2018think,
  title={Think you have solved question answering? try arc, the ai2 reasoning challenge},
  author={Clark, Peter and Cowhey, Isaac and Etzioni, Oren and Khot, Tushar and Sabharwal, Ashish and Schoenick, Carissa and Tafjord, Oyvind},
  journal={arXiv preprint arXiv:1803.05457},
  year={2018}
}

\end{document}